\documentstyle[epsf]{elsart}  
\begin{document}
\newcommand{\p}{\partial}
\newcommand{\ls}{\left(}
\newcommand{\rs}{\right)}
\newcommand{\beq}{\begin{equation}}
\newcommand{\eeq}{\end{equation}}
\newcommand{\beqa}{\begin{eqnarray}}
\newcommand{\eeqa}{\end{eqnarray}}
\newcommand{\bdm}{\begin{displaymath}}
\newcommand{\edm}{\end{displaymath}}
\begin{frontmatter}
\title{
Lambda collective flow in heavy ion reactions
} 
\author{
Z.S. Wang, Amand Faessler, C. Fuchs and T. Waindzoch
}
\address{Institut f\"ur Theoretische Physik der 
Universit\"at T\"ubingen, D-72076 T\"ubingen, Germany}
\begin{abstract}
Collective flow of Lambda hyperons in heavy ion reactions at
SIS energies is investigated. It is found that a $\Lambda$ mean field
constructed on the basis of the quark model leads to a good description of
the experimental data of the in-plane transverse flow of $\Lambda$'s.
The attractive
mean field can also give rise to an additional "virtual" $\Lambda$ radial flow
directed inwards, which is
reflected by a "concave" structure of the transverse mass spectrum
of the $\Lambda$ hyperons emitted at midrapidity. The $\Lambda$
radial flow is found to exhibit a strong  mass dependence:
The flow is visible in the Ni+Ni system, but is strongly reduced in
the system of Au on Au. 
\end{abstract}
\end{frontmatter}
\vskip 0.5 true cm
\section{Introduction}
Potentials of hyperons such as $\Lambda$, $\Sigma$, $\Xi$ etc.
in a nuclear environment have attracted interest
not only in nuclear physics, but also in astrophysics. It has long
been realized that hyperons will appear in the dense core of neutron
stars where the conversion of nucleons to hyperons through the
weak interaction is energetically favored because at the Fermi surface
the total energy of the nucleons will otherwise exceed the mass of
hyperons\cite{ambar60,glend91}. The presence of hyperons significantly softens the equation
of state of dense hadronic matter. This leads to the prediction of a lower
neutron star mass, which is closer to astrophysical measurements than
previous predictions for hyperon-free neutron stars\cite{glend91,balberg97}.
Other possible mechanisms to reduce neutron star masses include kaon
condensation which was suggested by Kaplan and Nelson\cite{kaplan86}.
Hyperon-nucleus potentials can be extracted from
hypernuclei\cite{batty97}. Theoretical models have also been used
to study hyperon-nucleus potentials. The approaches vary from
meson-exchange \cite{holze89,rufa90,glend93} to QCD (Quantum Chromodynamics)-based models\cite{jin94,saito97}.
According to the quark model, a hyperon interacts with nucleons
predominately through its non-strange quark content\cite{mosz74}. Consequently,
potentials of different hyperons in nuclear matter should satisfy
a simple scaling in terms of the number of non-strange quarks in
the hyperons\cite{saito97}. This speculation should be confronted with precise
experimental tests.
\vskip 0.5 true cm
Heavy ion reactions can serve as an alternative way for the investigation
of hyperon-nucleus potentials. In fact, heavy ion reactions provide
the only way to study in laboratories hyperon properties in
a nuclear environment which is much denser and hotter than
nuclei in their ground states. On the other hand, collective flow analysis
turns out to be powerful for
the study of the in-medium particle dynamics in heavy ion
reactions. Experimental data about the flow of nucleons
have been shown to yield stringent
constraints on the nuclear equation of state (EOS)\cite{pan93}. Recent
studies revealed that the flow of kaons provides valuable information on kaon
properties in the nuclear medium\cite{li95,wang97,wang98}.
In this paper we will study the flow of
Lambda hyperons in heavy ion reactions.

\vskip 0.5 true cm
We will explore in the present work two types of collective flows
of $\Lambda$'s, namely the in-plane transverse flow and the radial flow.
The transverse flow has been intensively studied for both
nuclear fragments and for secondary particles, while the radial
flow has been known conventionally as a collective
expansion of the dense and hot fireball formed in colliding nuclei.
In a recent work\cite{wang98}, we have shown that secondary particles
such as $K^+$ and
$K^-$ mesons can also develop a collective motion
in the same direction as the radial expansion of the fireball.
This novel collective
flow of the secondary particles is called "radial flow" as well.
The present work will investigate the possible formation of the
$\Lambda$ radial flow. We assume in the present study heavy ion beam
energies between 1 and 2 GeV/nucleon.
This energy
region is near the $\Lambda$ production threshold in free nucleon-nucleon
collisions (1.58GeV), and is well studied experimentally at Berkerley
and Darmstadt. This paper is organized as follows. In section 2, we
describe the $\Lambda$ dynamics in heavy ion reactions within the framework of
Quantum Molecular Dynamics (QMD). Section 3 is devoted to the
transverse flow of $\Lambda$'s. In section 4 we explore
the radial flow of $\Lambda$'s.
Section 5 is a summary of the paper.

\section{Lambda dynamics in the QMD model}
In heavy ion reactions at beam energies of 1-2 GeV/nucleon Lambda's
are produced mainly via the reactions B+B $\rightarrow$ B+$\Lambda$+$K^+$
and $\pi$+B $\rightarrow$ $\Lambda$+$K^+$\cite{fu97}. Here B stands for a nucleon or
a nucleonic resonance $\Delta$ or $N^*$. Due to strangeness conservation
a $\Lambda$ hyperon is accompanied by a positive charged kaon in both
reactions. So we treat the Lambda production in the QMD model in a similar
way as we did for $K^+$ mesons in a previous work\cite{wang97}.
We have used for both reactions calculated
cross sections which agree with the experimental data\cite{cassing97,tsu95}.
While the baryon-baryon (B-B) reaction is mainly isotropic,
the $\pi$-baryon ($\pi$-B) reaction
exhibits a strong asymmetric angular distribution due to the P-wave
contribution. The anisotropy of the $\pi$B reaction
has been taken into account with a parametrization of the experimental data.
This enabled
us to reproduce very well the experimental angular distribution of $K^+$'s 
in heavy ion reactions\cite{wang972}. A Lambda hyperon can not be absorbed
in nuclear matter. It, however, experiences strong
elastic scatterings on nucleons. The $\Lambda$N elastic cross section is
about 100 mb at a $\Lambda$ laboratory momentum of 0.2 GeV/c, and becomes
even larger with decreasing momenta\cite{baldini88}. We include Lambda-nucleon scatterings
with use of a parametrization of the experimental cross section\cite{li96}.

\vskip 0.5 true cm
The $\Lambda$ mean field in nuclear matter is constructed on the basis of
the quark model as usual\cite{mosz74,li96}.
The vector and scalar potentials acting on
a $\Lambda$ hyperon are taken to be two thirds of that experienced
by a nucleon, namely

\beq
\Sigma_S^{\Lambda} = \frac{2}{3}\Sigma_S^N
\eeq

\beq
\Sigma_V^{\Lambda} = \frac{2}{3}\Sigma_V^N
\eeq
  
where $\Sigma_S^N$ and $\Sigma_V^N$ are the nucleon scalar and vector
potentials evaluated in the non-linear $\sigma$-$\omega$ model\cite{boguta77}. The
mean-field potential of Lambda's is then defined as the difference
between the in-medium dispersion relation and the free one:
\beq
U_{\Lambda} = \sqrt{(m_{\Lambda} - \Sigma_S^{\Lambda})^2 + \vec{p}^2}
+ \Sigma_V^{\Lambda} -  \sqrt{m_{\Lambda}^2 + \vec{p}^2}
\eeq
In Fig.1 we present the $\Lambda$ potential at zero momentum as a function
of nuclear matter densities. This potential agrees at saturation density
$\rho_0$ = 0.16$fm^{-3}$ roughly with the value
extracted from hypernuclei\cite{millener88}. One can see from the figure
that the $\Lambda$ potential is attractive at nuclear matter densities
below three times the saturation density ($\rho$ $<$ 3$\rho_0$). The maximal
density reached in heavy ion reactions at 1-2 GeV/nucleon is about
3$\rho_0$. Therefore, one will see in these reactions mainly
effects of an attractive potential. We notice that the
$\Lambda$ potential (eq.3) is defined in a non-covariant way
since the contribution of the spatial components of the nucleon
current is neglected\cite{fuchs98}.
This approximation should
be justified for the present study due to twofold reasons. First,
the velocity of $\Lambda$'s is quite limited in the center-of-mass
frame of the colliding nuclei, since we consider in the present paper
a beam energy close to
the $\Lambda$ production threshold for free NN collisions. Second,
the $\Lambda$ hyperons experience frequent scatterings with nucleons,
which further reduce dramatically the $\Lambda$ velocity relative to 
surrounding nucleons. Therefore, the flow of the Lambdas should not
change by much as one changes from the non-covariant $\Lambda$
dynamics to a fully covariant one.
\vskip 0.5 true cm
The motion of $\Lambda$ hyperons is determined with the Hamiltonian:
\beq
H_{\Lambda} = \sum_{i_{\Lambda}}\left (\sqrt{m_{\Lambda}^2+{\vec P}_{i_{\Lambda}}^2}+
U_{i_{\Lambda}}\right ).
\eeq
\vskip 0.5 true cm
The $\Lambda$ production probability in an in-medium hadron-hadron
collision could be different from that of a collision in free space.
The magnitude of this medium effect depends on the in-medium potentials
of all the hadrons involved in the collision, i.e., nucleons, pions,
kaons and $\Lambda's$. The potentials of these hadrons have been not yet
well determined both experimentally and theoretically. Uncertainties
in evaluating the medium effect may also arise from the procedure to
decompose the interactions of the hadrons with the nuclear medium
into a long-range mean-field part and
a short-range hard-scattering part. The magnitude of the medium modification
of the $\Lambda$ and $K^+$ production probability
is still under current debates\cite{cassing97,ko96}. However, such
an effect should not change the collective flow of the $\Lambda$'s,
and therefore is neglected in the present work.
\vskip 0.5 true cm
\section{Lambda transverse flow}
Transverse flow is a collective emission of particles sidewards
in the reaction plane of colliding nuclei. This type of collective flow
can be quantitatively described
with the particle
momentum projected onto the reaction plane as a function of rapidity\cite{danie85}.
A flow results in a non-zero in-plane momentum at
projectile and target rapidity.
The existence of the transverse flow was originally
predicted for nuclear matter by the hydrodynamic
model\cite{glassgold59,stoecker80}, and was later verified by
experiments\cite{gustafsson84}.
It has been found that the transverse flow of nucleons or composite
particles such as deuterons, heliums etc. exhibits interesting
dependence on the bombarding energy. At low energies where the attractive
mean field dominates the interaction, the projectile fragments
are deflected towards the target\cite{tsang86}. If the bombarding
energy is high so that the overlapping matter is strongly compressed,
positive
pressure will develop in the interaction zone. Consequently, the
projectile fragments will be deflected away from the target
(so-called "bounce-off")\cite{gustafsson84}.
Between the two extreme cases, one finds a certain
incident energy where the attractive component and repulsive component of
the nuclear mean field counterbalance each other, and therefore
the transverse flow disappears\cite{krofc89,croch97,xu91}.
This special energy, called in literature "balance energy", is about
$E_{bal}$ = 10-25 MeV/nucleon as found experimentally\cite{krofc89,croch97}.
The beam energies considered in the present work (1-2 GeV/nucleon) are much
higher than $E_{bal}$. We therefore will see a bounce-off of nucleons for
the reactions under consideration\cite{wang97}.
\vskip 0.5 true cm
The studies of the transverse flow of nuclear fragments have motivated
a lot of recent studies
which analyzed the in-plane momentum of secondary particles such as
pions, kaons, antiprotons etc. at varying
rapidity\cite{li95,wang97,bli91,kint97}.
In analogue to the nucleon flow, one usually says that a transverse flow
of the secondary particles happens as a non-zero in-plane
momentum of the particles is observed at projectile or target rapidity.
The transverse flow of the secondary particles turns out to provide
information on the in-medium dynamics of these particles. The pion
transverse flow was found to be anticorrelated to the nucleon
flow\cite{bli91,kint97}. This behavior of the pion flow can be attributed
to rescatterings and reabsorptions of the pions in the nuclear
medium. The kaon transverse flow was shown to be sensitive to
in-medium modifications of kaon properties\cite{li95,wang97}. In this section
we explore the transverse flow of $\Lambda$'s, and its possible
sensitivity to the Lambda-nucleus potential.
\vskip 0.5 true cm
We present in Fig.2 the $\Lambda$ in-plane momentum as a function of rapidity
calculated from the QMD model for the reaction Ni+Ni at an incident energy
of 1.93 GeV/nucleon and within an impact parameter region of b = 0-4 fm.
Also shown in the figure are the experimental data of the FOPI Collaboration\cite{ritman95}.
A transverse momentum cut of $P_t$/$m_{\Lambda}$ $>$ 0.5 has been used
in obtaining both the experimental and the theoretical spectrum.
$\Lambda$ rapidity is calculated in the 
center-of-mass frame of the reaction system and normalized to the projectile
rapidity. We have performed the QMD calculations in two cases: with and without
the in-medium $\Lambda$ potential. One can see from the figure that
the calculation without the potential deviates from the experimental data at
high rapidity, while the full calculation reproduces well the data.
The FOPI experiment has found that the $\Lambda$ flow essentially follows
the proton flow in the same reaction, while the positive charged kaon flow is very
weak (The kaon in-plane momenta are nearly zero.). In a previous study
with the same QMD model as used in the current work, we have got
a very good agreement with the FOPI data for both proton and kaon flow\cite{wang97}.
As can be seen in Fig.2 the $\Lambda$ flow obtained without the potential
is weaker than the one from the full calculation. One can understand
the effect of the potential on the $\Lambda$ flow in a similar way
as one did for the kaon flow.
Provided that the Lambda's and kaons would experience neither
rescattering nor an in-medium potential, they
should have a same flow pattern, since
a $\Lambda$ hyperon and a $K^+$ meson are created simultaneously
from a BB or $\pi$B collision. The $K^+$ mesons feel
in the nuclear medium a repulsive potential which repels the kaons away from
their sources and by that leads to a near-vanishing $K^+$ transverse flow. 
$\Lambda$ hyperons, on the contrary, experience an attractive mean field
at the beam energies under consideration. Thus the $\Lambda$ flow
changes to be more pronounced due to the potential.
\vskip 0.5 true cm
Aside from the $\Lambda$ mean field, $\Lambda$-nucleon scatterings
also affect the $\Lambda$ flow. At the beam energy under consideration
the projectile-like and the target-like remnants are deflected
away from each other as mentioned above. The transverse motion of these remnants
can transfer to the Lambda's through the $\Lambda$-nucleon scatterings.
Consequently, the $\Lambda$ flow is enhanced due to the $\Lambda$-nucleon
scatterings.
\vskip 0.5 true cm
The $\Lambda$ transverse flow of the same reaction has also been studied
in Ref.\cite{li96} with a different transport model, i.e. the Relativistic
Boltzmann-Uehling-Uhlenbeck (RBUU) model. Authors of Ref.\cite{li96}
used a non-covariant description of the $Lambda$ dynamics which is quite
similar to the one adopted by the present paper. Good agreement
is found between Ref.\cite{li96} and the present study concerning the
transverse flow of $K^+$'s and $\Lambda$'s. As far as the proton flow
is concerned, small discrepancies exist between the two theoretical studies.
Our QMD model turns out to reproduce very well the proton flow
data up to $y_{cm}/y_{proj} = 1.5$, while the RBUU model of Ref.\cite{li96}
slightly underpredicts the data at $y_{cm}/y_{proj}$ $>$1.
\vskip 0.5 true cm
\section{Lambda radial flow}
Radial flow conventionally means a
subsequent expansion of the dense and hot fireball formed
in heavy ion reactions\cite{siemens79,lisa95}.
The fireball expansion has similar origin as the bounce-off: both
are driven by the positive pressure in the overlap region
of colliding nuclei. The threshold beam energy for onset of the
expansion is found experimentally to be about 35
GeV/nucleon\cite{gobbi95}, slightly higher than the threshold
energy for the occurrence of the bounce-off phenomenon, $E_{bal}$.
The fireball expansion
gives rise to some non-thermal behavior of nucleons\cite{siemens79},
composite particles such as deuterons, tritons etc.\cite{mattiello95}.
It can also
implement non-thermal features to secondary particles, provided
that the beam energies are so high that hadron-hadron collisions in the late
expansion phase of the reaction are still energetic enough to produce these
particles. This condition is fulfilled at AGS and SPS energies
( about 10 GeV/nucleon and 150 GeV/nucleon, respectively) for pions,
kaons, Lambda's etc.. Therefore, one can find information on the fireball
expansion not only from nucleon observables but also from observables
of these secondary particles\cite{bear97,schne92,ahmad98,danie95}.
At a subthreshold beam energy,
one will hardly find effects of the fireball
expansion on secondary particles, since the particles are produced
mainly in the compression stage prior to the subsequent fireball expansion.
In a recent study,
we demonstrated that $K^+$ and $K^-$ mesons produced at a subthreshold
beam energy (1.58 GeV/nucleon for $K^+$'s and 2.5 GeV/nucleon for $K^-$'s)
can exhibit
non-thermal features, however, due to their in-medium potentials instead of
the fireball expansion\cite{wang98}. $K^+$ ( $K^-$ ) mesons feel a repulsive
(attractive) potential in nuclear matter, which mutually accelerates (decelerates)
the $K^+$ ($K^-$) mesons as they escape from the dense fireball. This
results in a "shoulder-arm" ("concave") structure in the transverse
mass spectrum of the $K^+$ ($K^-$) mesons emitted at midrapidity, which
obviously deviates from a Boltzmann distribution. We called kaon radial flow
the common acceleration or deceleration by the in-medium potentials.
\vskip 0.5 true cm
In this section we investigate if $\Lambda$ hyperons can also
develop a radial flow similar to kaons. Fig.3 shows the transverse mass spectrum
of the $\Lambda$ hyperons emitted at midrapidity ( -0.4$<$$y_{cm}$/$y_{proj}$$<$
0.4 ) calculated for the same reaction as in Fig.2. A "concave" structure can be clearly
observed in the spectrum at small values of the transverse mass
($m_T$- $M_{\Lambda}$ = $\sqrt{m_{\Lambda}^2+p_T^2}$-$m_{\Lambda}
$ $<$ 0.2 GeV). This feature
obviously distinguishes the $\Lambda$ spectrum from a Boltzmann distribution
(see a Boltzmann fit to the high-energy part of the $\Lambda$ spectrum
which is also shown in Fig.3). The "concave" structure of the $\Lambda$
spectrum arises from the attractive $\Lambda$ mean field. If the $\Lambda$
potential is neglected, the "concave" shape disappears. This can be seen in
Fig.4 where the $\Lambda$ transverse mass spectrum from
the QMD calculation without the potential is presented. The $\Lambda$ spectrum
is now exactly a Boltzmann distribution.
\vskip 0.5 true cm
In order to extract information on the in-medium potentials from the
transverse mass spectrum, we suggested in Ref.\cite{wang98} an equation
to fit the spectrum. The fit function
has been derived by boosting a thermal Boltzmann
distribution with a common radial velocity due to the in-medium potential.
\beq
\frac{{d^3}N}{d{\phi}dy{m_t}d{m_t}} \sim e^{-(\frac{{\gamma}E}{T}+{\alpha})}
\{ {\gamma}^{2}E + {\gamma}{\alpha}T(\frac{E^2}{p^2}+1) + ({\alpha}T)^{2}\frac{
E^2}{p^2} \}\frac{\sqrt{({\gamma}E+{\alpha}T)^{2}-m^{2}}}{p}
\label{reswidth}
\eeq
where $E$ = $m_t$$coshy$, $p$ = $\sqrt{{p_t}^2+{m_t}^2sinh^2y}$,
$\alpha$ = $\gamma$$\beta$$p/T$, $\gamma$ = $(1-\beta^2)$$^{-1/2}$.
There are two free parameters appearing in the fit equation (eq.5).
Both of them have clear physical meanings:
T is the average temperature of $\Lambda$ sources,
while the parameter $\beta$ = $\upsilon$/c is a common
radial velocity of the Lambda's caused by the potential. An attractive
potential leads to a reduction of $\Lambda$ radial velocities. We thus should
use in the fit equation a $\beta$ directed towards the center of the fireball.
One can extract the T value from the $\Lambda$ spectrum which comes from
the QMD calculation without the $\Lambda$ potential. This yields a value
of T = 115 MeV. The $\beta$ value can be evaluated by comparing the simplified
QMD calculation with the full one including the potential. We found from
the calculations that the average mass of the midrapidity Lambda's is
reduced by the attractive potential from $<m_T>$-$m_{\Lambda}$ = 127 MeV
to $<m_T>$-$m_{\Lambda}$
= 115 MeV. This reduction accounts for a $\beta$ value of $\beta$ = 0.025 in
the fit procedure. In Fig.3 we plot the result of the fit by using eq.5
with T and $\beta$ parameters determined as above. A satisfactory agreement
can be found from the figure between the full QMD calculation and the
fit procedure. This demonstrates that a $\Lambda$ radial flow forms
in the reaction due to the attractive potential. However, the energy
reduction caused by the potential is rather small compared to the thermal
energy: the former is only about 10$\%$ of the latter. Therefore, experimental
data of high accuracy are required in order to identify the $\Lambda$
radial flow. We have noticed that the EOS Collaboration reported a measurement
of the transverse mass spectrum of midrapidity $\Lambda$'s in
the reaction Ni + Cu at 2 GeV/nucleon\cite{eos97}. However, the error bars of the
EOS data are too large (in particular in the region of 0.2 GeV$<$
$m_T$-$m_{\Lambda}$ $<$ 0.5 GeV) to enable us to draw any conclusion
about the $\Lambda$ radial flow.
\vskip 0.5 true cm
It is interesting to study the influence of the size of the reaction
system on the $\Lambda$ radial flow. In order to do this, we also study
with the QMD model reactions induced by massive nuclei, i.e. Au+Au.
Fig.5 shows
the calculated transverse mass spectrum of midrapidity Lambda's for this reaction
at an incident energy of 1 GeV/nucleon and at an impact parameter of
b = 3 fm. This spectrum can be described very well by a pure Boltzmann
distribution as can be seen in the figure. In Fig.6 we present
the $\Lambda$ spectrum of the same reaction as in Fig.5 from a calculation where the $\Lambda$ mean field
is neglected. One finds again that the QMD result without the $\Lambda$
potential agrees well with a thermal distribution. The $\Lambda$
spectrum from the simplified calculation is very similar to the one
from the full calculation (Fig.5). The average transverse mass $<m_t>$
is only reduced by about 1 MeV due to the potential. Thus, no effect
of the $\Lambda$ mean field can be observed and the radial flow is invisible
in the Au+Au reaction.
\vskip 0.5 true cm
That $\Lambda$ radial flow forms in the Ni+Ni reaction but disappears
in the Au+Au reaction can be attributed to $\Lambda$-nucleon scatterings.
It is clear from Fig.4 and Fig.6 that $\Lambda$-nucleon scatterings
alone can not induce any two-temperature structure of the $\Lambda$
spectrum. However,
as soon as a $\Lambda$ radial flow develops as a consequence of
the attractive potential, $\Lambda$-nucleon scatterings begin to play
a role in canceling the effect of the potential, since the scatterings
bring the $\Lambda$'s from the low energy region of $m_T$ - $m_{\Lambda}$
$<$ 0.2 GeV to higher energies.
As we have
mentioned, a $\Lambda$ hyperon has a short mean free path in nuclear
matter, especially for low-energy Lambda's. One expects more
$\Lambda$-nucleon scatterings in a larger system,
and thus less low-$m_T$ Lambda's. Consequently,
one finds no "concave" structure in the $\Lambda$ spectrum
in the reaction Au+Au, while
this structure survives in the Ni+Ni reaction as a result of
reduced $\Lambda$-nucleon scatterings. The density dependence of
the $\Lambda$ potential could also play a role in
the $\Lambda$ radial flow. As can be seen in Fig.1,
the $\Lambda$ potential changes to be shallower as the density
increases beyond the normal nuclear matter density $\rho_0$. A larger system
has a larger stopping power, and thus a longer lifetime of the dense
fireball. Consequently, the $\Lambda$ hyperons experience a less attractive
potential in a larger system. This also leads to a weaker radial flow
in the Au+Au reaction than in Ni+Ni reaction.
\vskip 0.5 true cm
It is worthwhile to indicate that the $\Lambda$ radial flow is
very sensitive to incident energies. The conclusions of the present work
are drawn for the incident energies close to the $\Lambda$
production threshold for free NN collisions.
These relatively low beam energies, i.e. 1-2 GeV/nucleon, lead to
a negligible effect of the fireball expansion on the $\Lambda$'s
as we have mentioned. However, as one goes to AGS energies or even
SPS energies, the fireball expansion will play a non trivial role
in determining the $\Lambda$ spectrum. The fireball
expansion can reduce the multiplicity of low-$m_T$ $\Lambda$'s by
pushing them to a finite $m_T$ corresponding to the common expansion
velocity of the fireball. Consequently, the transverse mass spectrum
of the $\Lambda$'s
will show a "shoulder-arm" shape rather than a "concave"
shape. Such "Shoulder-arm" structure has been observed by the E891
Collaboration for the incident energy of 11.6 GeV/nucleon\cite{ahmad98}. This
observation does not contradict with the present study.

\vskip 0.5 true cm
\section{Summary}
In this paper we have studied the collective flow of $\Lambda$ hyperons in
heavy ion reactions at SIS energies with the QMD model. It is shown
that the FOPI data of the $\Lambda$ transverse flow,
which is quantitatively
described with the $\Lambda$ in-plane momentum as a function of rapidity,
can be well reproduced by using a $\Lambda$ potential constructed based on
the quark model. The QMD model used in the present paper can also
describe very well the measured proton and kaon flow as shown in
our previous work. A radial flow of $\Lambda$ hyperons is found to arise
from the attractive $\Lambda$ potential in the Ni+Ni reaction. The flow results in a "concave" structure
of the transverse mass spectrum of the $\Lambda$ hyperons emitted
at midrapidity. However, as one changes to the reaction
Au+Au, the $\Lambda$ radial flow vanishes.
The size dependence of the $\Lambda$ radial flow can be attributed to
frequent $\Lambda$-nucleon scatterings and reduced attraction of
the $\Lambda$ potential in a large system.

\newpage
Figure Captions
\vskip 0.5 true cm

Fig. 1. The $\Lambda$ mean-field potential
in nuclear matter used in this work (at zero momentum). The
circle denotes the potential at normal nuclear matter density
$\rho_0$ extracted from hypernuclei experiments\cite{millener88}.

\vskip 0.5 true cm

Fig. 2. The in-plane $\Lambda$ momentum as a function of rapidity
from the Ni+Ni reaction at 1.93 GeV/nucleon and within an impact parameter
region of b=0-4 fm. The solid line denotes the QMD calculation
including the full in-medium Lambda dynamics, while the dashed line
is the QMD calculation without the $\Lambda$ potential. The experimental
data from the FOPI Collaboration are also presented. 
\vskip 0.5 true cm

Fig. 3. The transverse mass spectrum of the $\Lambda$ hyperons emitted at
midrapidity ( -0.4$<$$y_{c.m.}$/$y_{proj}$$<$0.4 ) from the same reaction
as in Fig.2.
The histogram presents the result of the
QMD calculation
which is performed including the full in-medium Lambda dynamics.
The dashed line presents the result of a pure thermal Boltzmann
fit to the high energy part of the
spectrum ( $m_t$-$m_0$ $>$ 0.2 GeV ), while the dotted line
is the fit according to eq.(5) where one assumes a $\Lambda$ radial
flow in addition to the thermal motion.

\vskip 0.5 true cm
Fig. 4. The transverse mass spectrum of the $\Lambda$ hyperons emitted at
midrapidity ( -0.4$<$$y_{c.m.}$/$y_{proj}$$<$0.4 ) from the same
reactions as in Fig. 2. The histogram presents the result of the
QMD calculation without the $\Lambda$ mean field.
Now a pure thermal fit (lines )
is sufficient to reproduce the spectrum.

\vskip 0.5 true cm
Fig. 5. The transverse mass spectrum of the $\Lambda$ hyperons emitted at
midrapidity ( -0.4$<$$y_{c.m.}$/$y_{proj}$$<$0.4 ) from the
reaction Au+ Au at 1 GeV/nucleon and b=3 fm. The histogram presents the result of the
QMD calculation with the $\Lambda$ mean field, while
the line is a Boltzmann fit to the QMD result.

\vskip 0.5 true cm
Fig. 6. The transverse mass spectrum of the $\Lambda$ hyperons emitted at
midrapidity ( -0.4$<$$y_{c.m.}$/$y_{proj}$$<$0.4 ) from the same
reaction as in Fig.5. The histogram presents the result of the
QMD calculation without the $\Lambda$ mean field, while
the line is a Boltzmann fit to the QMD result.
\newpage
\begin{figure}
\leavevmode
\epsfxsize = 12cm
\epsffile[60 85 430 750]{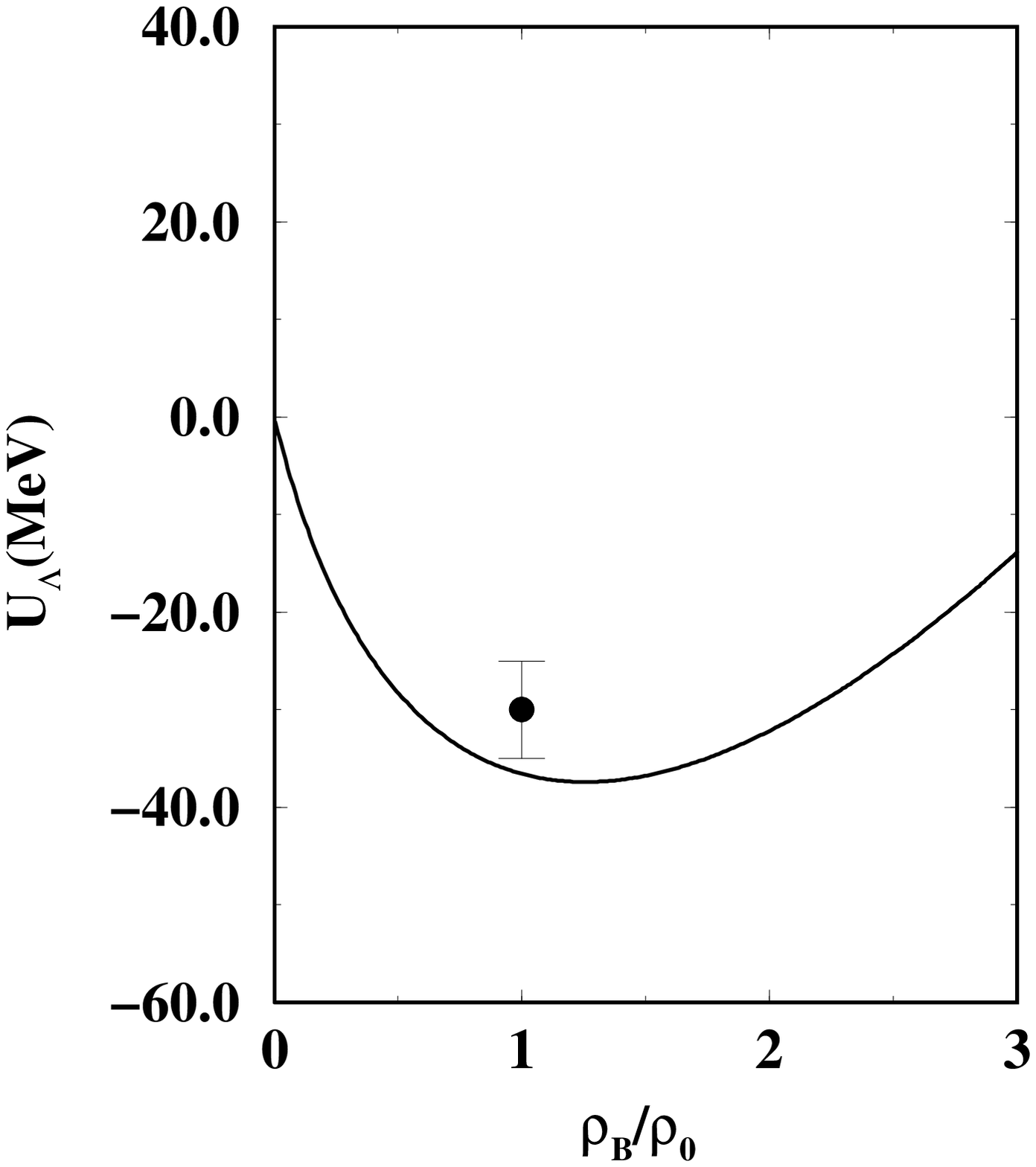}
\caption{
}
\label{fig2}
\end{figure}
\newpage
\begin{figure}
\leavevmode
\epsfxsize = 12cm
\epsffile[60 85 430 750]{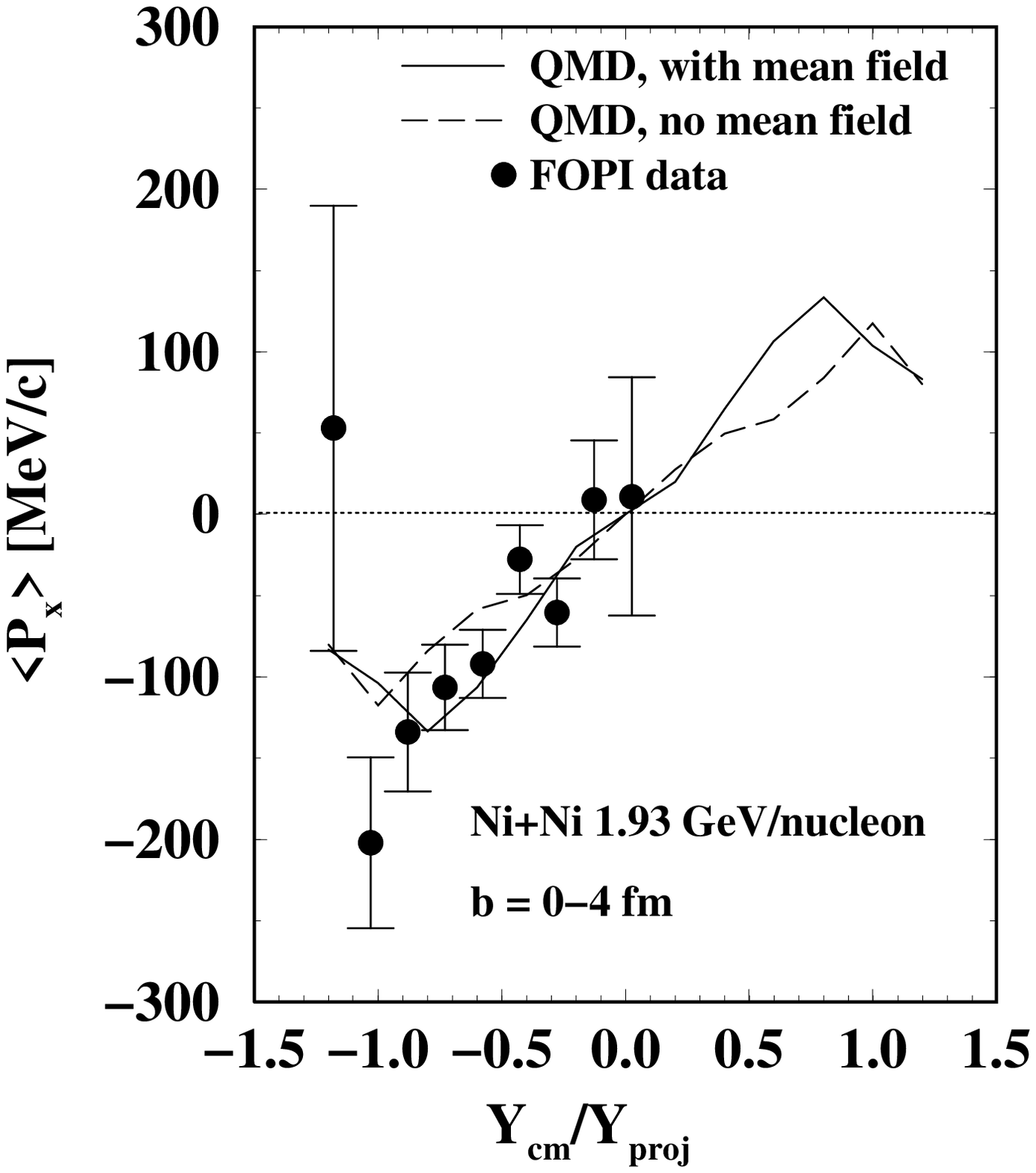}
\caption{
}
\label{fig2}
\end{figure}
\newpage
\begin{figure}
\leavevmode
\epsfxsize = 12cm
\epsffile[60 85 430 750]{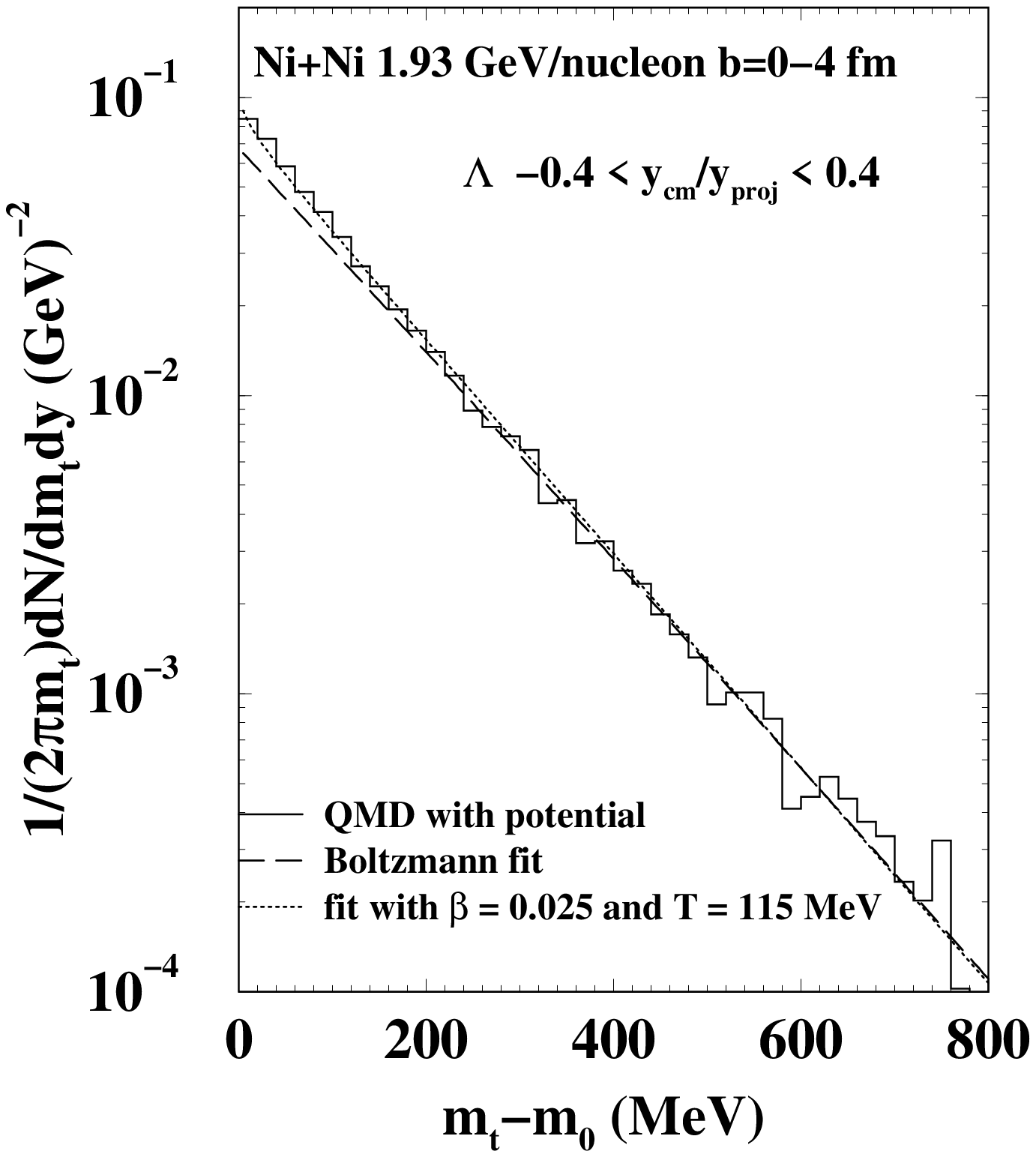}
\caption{
}
\label{fig3}
\end{figure}

\newpage
\begin{figure}
\leavevmode
\epsfxsize = 12cm
\epsffile[60 85 430 750]{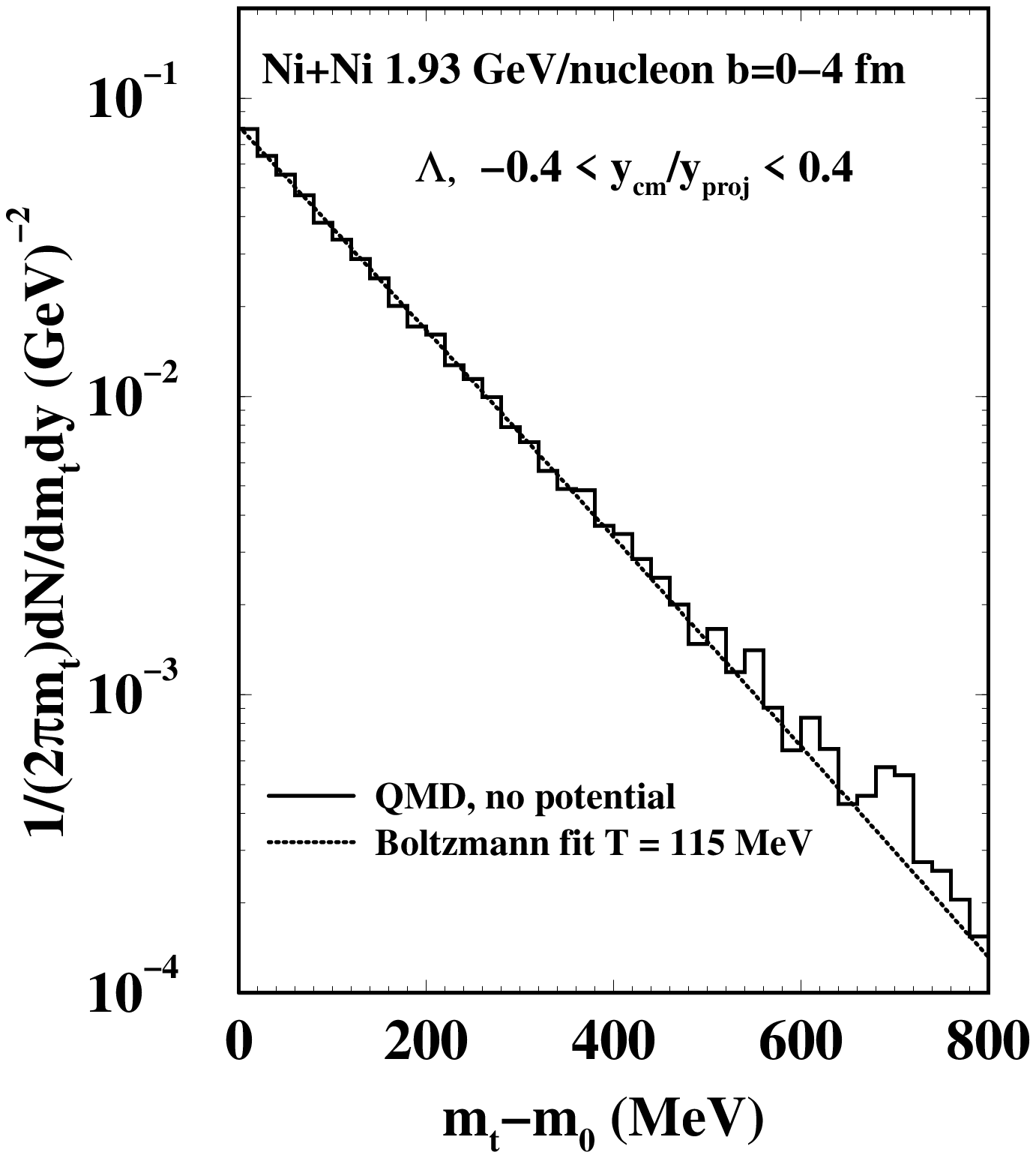}
\caption{
}
\label{fig4}
\end{figure}

\newpage
\begin{figure}
\leavevmode
\epsfxsize = 12cm
\epsffile[60 85 430 750]{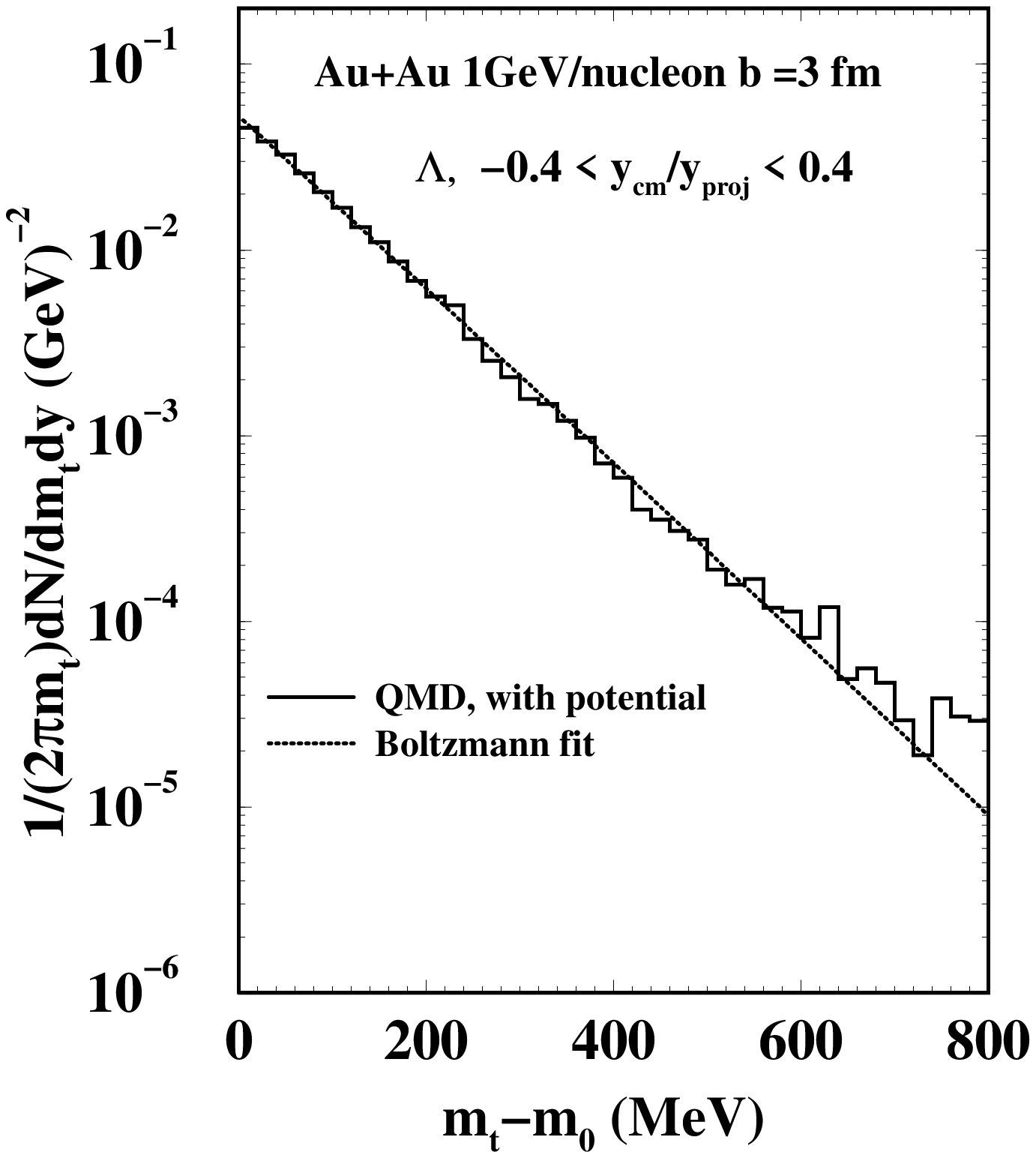}
\caption{
}
\label{fig5}
\end{figure}

\newpage
\begin{figure}
\leavevmode
\epsfxsize = 12cm
\epsffile[60 85 430 750]{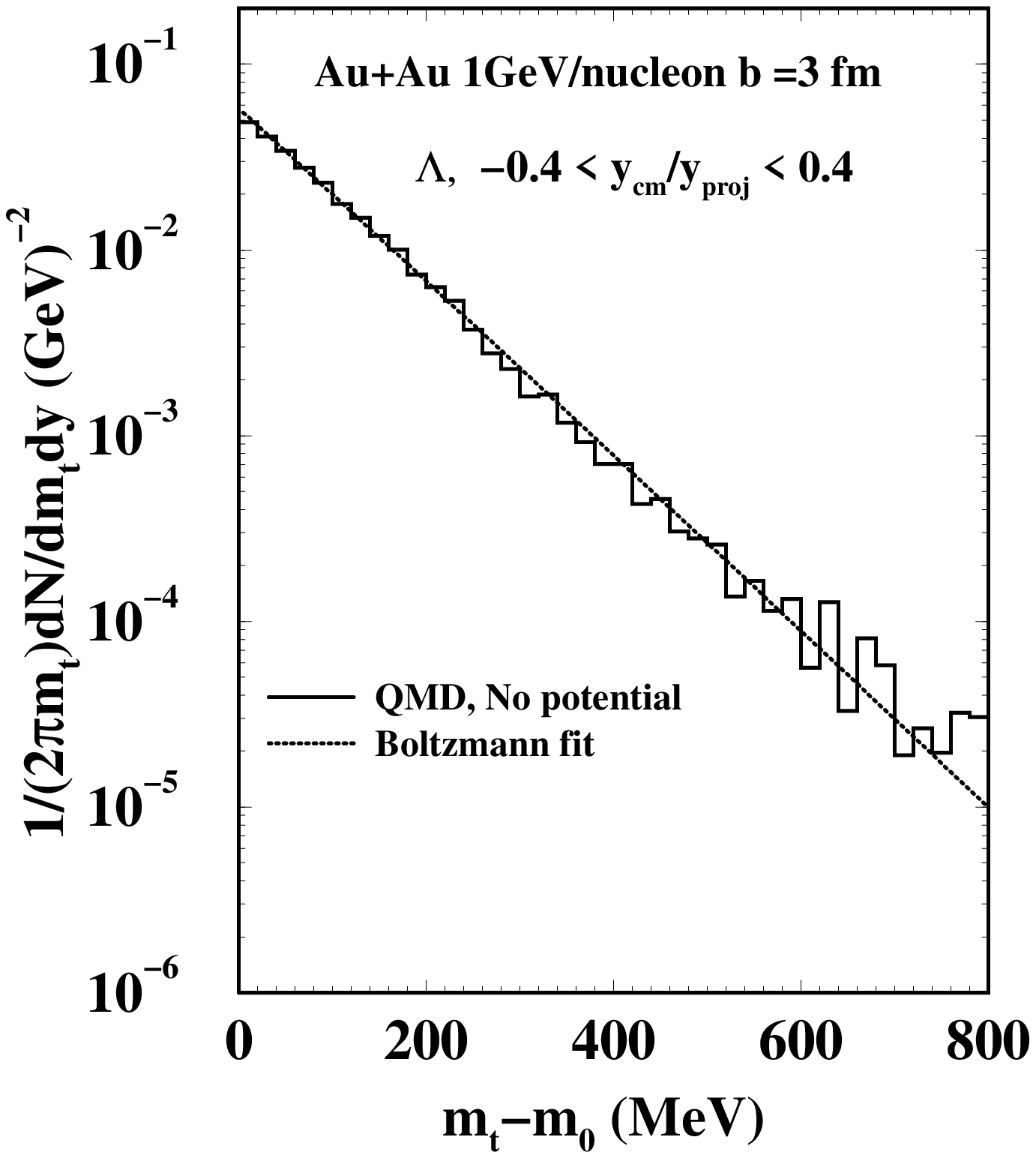}
\caption{
}
\label{fig6}
\end{figure}
\end{document}